\documentstyle[aps,prl,epsf,amssymb,amsmath,amsthm]{revtex}
\bibstyle{unsrt}

\tighten
\begin{document}
\draft \preprint{}


\title{All Hermitian Hamiltonians Have Parity}

\author{Carl M. Bender, Peter N. Meisinger, and Qinghai Wang}

\address{Department of Physics, Washington University, St. Louis, MO 63130, USA}

\date{\today}
\maketitle

\begin{abstract}
It is shown that if a Hamiltonian $H$ is Hermitian, then there always exists an
operator ${\cal P}$ having the following properties: (i) ${\cal P}$ is linear
and Hermitian; (ii) ${\cal P}$ commutes with $H$; (iii) ${\cal P}^2=1$; (iv) the
$n$th eigenstate of $H$ is also an eigenstate of ${\cal P}$ with eigenvalue
$(-1)^n$. Given these properties, it is appropriate to refer to ${\cal P}$ as
the parity operator and to say that $H$ has parity symmetry, even though ${\cal
P}$ may not refer to spatial reflection. Thus, if the Hamiltonian has the form
$H=p^2+V(x)$, where $V(x)$ is real (so that $H$ possesses time-reversal
symmetry), then it immediately follows that $H$ has ${\cal PT}$ symmetry. This
shows that ${\cal PT}$ symmetry is a generalization of Hermiticity: All
Hermitian Hamiltonians of the form $H=p^2+V(x)$ have ${\cal PT}$ symmetry, but
not all ${\cal PT}$-symmetric Hamiltonians of this form are Hermitian.
\end{abstract}

\vskip2pc

The requirement that a Hamiltonian be Hermitian guarantees that the energy
eigenvalues of the Hamiltonian are real. However, in 1998 \cite{BB} it was
shown that a non-Hermitian Hamiltonian can still have an entirely real spectrum
provided that it possesses ${\cal PT}$ symmetry. For example, with properly
defined boundary conditions, the Sturm-Liouville differential equation
eigenvalue problem associated with the non-Hermitian Hamiltonian
\begin{equation}
H=p^2+x^2(ix)^\nu\qquad(\nu>0)
\label{e1}
\end{equation}
exhibits a spectrum that is {\em real and positive}. It was argued in
Ref.~\cite{BB} that the reality of the spectrum of $H$ is a consequence of the
unbroken ${\cal PT}$ symmetry of $H$. A complete proof that the spectrum of $H$
is real and positive was given by Dorey et {\em al} \cite{DDT}.

In Ref.~\cite{BB} it was stated that ${\cal PT}$ symmetry (space-time reflection
symmetry) is a weaker condition than Hermiticity in the following sense. For
many different Hermitian Hamiltonians, such as $H=p^2+x^4$, $H=p^2+x^6$,
$H=p^2+x^8$, and so on, we can construct infinite classes of non-Hermitian
${\cal PT}$-symmetric Hamiltonians $H=p^2+x^4 (ix)^\nu$, $H=p^2+x^6(ix)^\nu$,
$H=p^2+x^8(ix)^\nu$, and so on. So long as the parameter $\nu$ is real and
positive ($\nu>0$), the ${\cal PT}$ symmetry of each of these Hamiltonians is
not spontaneously broken and the spectrum is entirely real \cite{BBM}.

In this paper we show that for Hamiltonians of the form $H=p^2+V(x)$, ${\cal
PT}$ symmetry is a generalization of Hermiticity and that the set of Hermitian
Hamiltonians (for which $V(x)$ is real) is {\em entirely contained} within the
set of ${\cal PT}$-symmetric Hamiltonians. That is, we will show that if a
Hamiltonian of this type is Hermitian, then it possesses both parity symmetry
${\cal P}$ and time-reversal symmetry ${\cal T}$.

As an example, consider the Hermitian Hamiltonian
\begin{equation}
H=p^2+x^4+x^3.
\label{e2}
\end{equation}
It is obvious that this Hamiltonian is symmetric under the operation of time
reversal ${\cal T}$, where ${\cal T}$ transforms $p\to -p$ and $x\to x$. (The
operator ${\cal T}$ also transforms $i\to-i$, but because the Hamiltonian is
real, this fact is not relevant here.) The Hamiltonian $H$ also possesses
another discrete symmetry that can be called parity. The purpose of this paper
is to show how to construct such an operator for any Hermitian Hamiltonian.

Given a Hamiltonian like that in (\ref{e2}) one may in principle solve the
time-independent Schr\"odinger equation 
\begin{equation}
H\phi_n(x) = E_n \phi_n(x),
\label{e3}
\end{equation}
where the eigenfunctions of $H$ are $\phi_n(x)$ and the corresponding
eigenvalues are $E_n$. These eigenfunctions form an orthonormal set:
\begin{equation}
\int dx\phi_m^*(x)\phi_n(x) = \delta_{m,n}.
\label{e4}
\end{equation}

From the theory of Hermitian operators we know that
the eigenfunctions form a complete basis:
\begin{equation}
\delta(x-y) = \sum_{n=0}^\infty \phi_n(x) \phi_n^*(y).
\label{e5}
\end{equation}
Because the coordinate-space eigenfunctions are complete we can use them to
represent the Hamiltonian as a matrix in coordinate space:
\begin{equation}
H(x,y)=\sum_{n=0}^\infty E_n\phi_n(x)\phi_n^*(y).
\label{e6}
\end{equation}

Let us now follow the approach of Ref.~\cite{BBJ} to construct a new operator,
which we will call ${\cal P}(x,y)$:
\begin{equation}
{\cal P}(x,y) \equiv \sum_{n=0}^\infty (-1)^n \phi_n(x) \phi_n^*(y).
\label{e7}
\end{equation}
Observe that ${\cal P}$ has the following four properties:
(i) The operator ${\cal P}$ is linear and Hermitian.
(ii) ${\cal P}$ commutes with the Hamiltonian.
(iii) ${\cal P}^2 = 1$; that is, in coordinate space $\int dz\,{\cal P}(x,z)
{\cal P}(z,y) = \delta(x-y)$. (iv) $\phi_n$ is an eigenfunction of ${\cal P}$
with eigenvalue $(-1)^n$; that is,
\begin{equation}
\int dz {\cal P}(x,z) \phi_n(z)=(-1)^n \phi_n(x),
\label{e8}
\end{equation}
by virtue of orthonormality.

Based on property (i) the operator ${\cal P}$ is an observable, and based on
property (ii) this observable is conserved (time-independent). Moreover, because
of properties (iii) and (iv) the operator ${\cal P}$ exhibits the
characteristics of the parity operator even though the Hamiltonian may not be
symmetric under space reflection. We remark that if the potential $V(x)$ of the
Hamiltonian $H=p^2+V(x)$ is invariant under the transformation $x\to-x$, then
the operator ${\cal P}(x,y)$ in (\ref{e7}) is just the usual parity operator
$\delta(x+y)$. Note that ${\cal PT}$-symmetric Hamiltonians, for which ${\cal
P}$ is a more general symmetry than space reflection are considered in
Ref.~\cite{BBA}.

We can follow this procedure to construct many different operators that commute
with the Hamiltonian. For example, we can construct a ``triparity'' operator
${\cal Q}(x,y)$, whose cube is unity:
\begin{equation}
{\cal Q}(x,y) = \sum_{n=0}^\infty \omega^n \phi_n(x) \phi_n^*(y),
\label{e9}
\end{equation}
where $\omega=e^{\pm 2i\pi/3}$, so that $\omega^3=1$. However, this operator
is not an observable because it is not Hermitian.

We have shown that if a Hamiltonian of the form $H=p^2+V(x)$ is Hermitian then
it is also ${\cal PT}$ symmetric. (The converse is of course not true.) Thus,
${\cal PT}$ symmetry is demonstrated to be a generalization of Hermiticity.

\vskip2pc This work was supported by the U.S.~Department of Energy.

\end{document}